\documentclass{elsart}
\usepackage{epsfig}
\usepackage{graphicx}

\begin{document}
\begin{frontmatter}

\title{Dynamical random multiplicative cascade model in 1+1 dimensions}

\author{
J\"urgen  Schmiegel$^{a,b}$,
Hans C.\ Eggers$^{a}$,
and 
Martin Greiner$^{b,c,d}$
}

\address{$^{a}$Department of Physics, University of Stellenbosch,
               7600 Stellenbosch, South Africa }
\address{$^{b}$Max-Planck-Institut f\"ur Physik komplexer Systeme, 
               N\"othnitzer Str.\ 38, D--01187 Dresden, Germany }
\address{$^{c}$Department of Physics, Duke University, 
               Durham, NC 27708, USA } 
\address{$^{d}$Fysisk Institutt, Universitetet i Bergen,
               Allegaten 55, N-5007 Bergen, Norway } 

\begin{abstract}
Geometrical random multiplicative cascade processes are often used to
model positive-valued multifractal fields such as for example the 
energy dissipation field of fully developed turbulence. A dynamical 
generalisation of these models is proposed, which describes the 
continuous and homogeneous stochastic evolution of the field in one
space and one time dimension. Two-point correlation functions are
calculated.
\end{abstract}

\end{frontmatter}

\newpage
Whenever strong anomalous, intermittent fluctuations, long-range 
correlations, multi-scale structuring and selfsimilarity go hand in 
hand, the label `multifractality' is attached to the underlying process.
Although we have fully developed turbulence of fluid mechanics \cite{FRI95}
in mind, such processes occur in various diverse fields such as 
formation of cloud and rain fields of geophysics \cite{SCH87}, 
internet traffic of signal and communication engineering \cite{RIE99}, 
and stock returns of econometrics \cite{MUZ00}, just to name a few.
The maybe most widely used textbook \cite{FED88} visualisation for 
multifractality is represented by a multiplicative cascade process. It 
reproduces the above mentioned properties by introducing a 
scale-independent cascade generator, which produces a nested
hierarchy of scales and redistributes the local mass multiplicatively. 

In fully developed turbulence, random multiplicative cascade models (RMCM) 
are often employed to model the energy cascade, describing the energy flux 
through inertial range scales. Due to their multiplicative nature, it is 
straightforward for them to reproduce multifractal scaling exponents 
associated with the energy dissipation field \cite{MEN91}, the latter 
representing the intermittency corrections to the K41-theory \cite{FRI95}.
Although the justification of RMCMs in terms of the Navier-Stokes equation
is far from being clear, these phenomenological models appear to contain 
more truth than originally anticipated, as recent investigations on 
multiplier distributions \cite{JOU99} and scale correlations \cite{CLE00} 
have revealed. 
--
On the other side RMCMs are purely geometrical in nature and are not 
capable of describing causal dynamical effects of the turbulent energy
cascade like backscattering of energy flux from small to larger scales.
A generalisation in this direction is clearly called for.

In this Letter we present such a generalisation and construct a dynamical
RMCM in $1{+}1$ space-time dimensions, respecting causality. 
Multifractal scaling is recovered for $n$-point correlation functions. 

In a discrete RMCM in $1{+}0$ dimensions, the amplitude
\begin{equation}
\label{one}
  \varepsilon(\eta)
    =  \prod_{j=1}^{J} q(l_j)
    =  \exp\left( \sum_{j=1}^{J} \ln{q(l_j)} \right)
\end{equation}
of the positive-valued energy-dissipation field, resolved at the
dissipation scale $\eta$, is equal to the product of independently and
identically distributed random weights $q(l_j)$, which come with
$\langle q \rangle = 1$ and a nested hierarchy of scales
$\eta=l_J \leq l_j=L/\lambda^j \leq l_0=L$ with $0{\leq}j{\leq}J$. 
The integral length L and the dissipation length $\eta$ represent the 
largest and smallest length scale involved, whereas $\lambda{>}1$ is the 
discrete scale step. 

It is the second step of (\ref{one}), which we now exploit to propose 
a generalisation of the cascade field, now defined in $1{+}1$ 
space-time dimensions:
\begin{equation}
\label{two}
  \varepsilon(x,t)
    =  \exp\left\{ 
       \int_{-\infty}^{\infty}dt' \int_{-\infty}^{\infty}dx' 
       f(x-x',t-t') \gamma(x',t')
       \right\}
       \; .
\end{equation}
Due to the point-resolution the function 
$\gamma(x,t) \sim S_\alpha((dxdt)^{\alpha^{-1}-1}\sigma,-1,\mu)$ 
is taken as a stable white-noise field with index $0{\leq}\alpha{\leq}2$ 
\cite{SAM94}; for the special case of $\alpha{=}2$ it corresponds to a 
non-centred Gaussian white-noise field. Its characteristic function is  
$\langle\exp\{n\gamma\}\rangle
 = \exp\{-(\sigma^\alpha n^\alpha)/(\cos(\pi\alpha/2)) + \mu n\}$,
which fixes the parameter $\mu=\sigma^\alpha/\cos(\pi\alpha/2)$ in 
order to fulfill the expectation $\langle\exp\{\gamma\}\rangle=1$.
Imposing causality, i.e.\ the field amplitude $\varepsilon(x,t)$ is only
influenced by the past and not by the future, requires $f(x,t)=0$ for 
$t<0$. This is satisfied with the following symmetric index function
\begin{equation}
\label{three} 
  f(x,t)
    =  \left\{ \begin{array}{ll} 
       1 & \qquad (\,0{\leq}t{\leq}T, \; -g(t){\leq}x{\leq}g(t) \,) \\
       0 & \qquad (\mbox{otherwise}) \; ,
       \end{array} \right. 
\end{equation}
where the window function $g(t)$ introduces a correlation time $T$ and a
correlation length $L$ with $g(0) \approx 0$ and $g(T)=L/2$; see Fig.\ 1
for an illustration. The exponent of the construction (\ref{two}) can
be thought of as a moving average over the stable white-noise field.

According to (\ref{three}), the time integration in (\ref{two})
goes from $-T{\leq}t^\prime{\leq}0$, where $x=t=0$ has been set 
for simplicity. Upon assigning $2g(-t_j){=}l_j$ the hierarchy of
length scales $l_j$ is translated into a hierarchy of time scales, so that
(\ref{two}) then factorises into contributions
\begin{equation}
\label{four}
  q(l_j)
    =  \exp\left\{
       \int_{t_{j-1}}^{t_{j}}dt' 
       \int_{-g(-t')}^{g(-t')}dx' \gamma(x',t')
       \right\}
       \; ;
\end{equation}
see again Fig.\ 1.
In order to interpret this as a random multiplicative weight,
the probability distribution density of $q(l_j)$ needs to be 
independent of scale. Since the $\gamma(x',t')$ are i.i.d., the
integration domain of (\ref{four}) then has to be independent of
the scale index $j$. This fixes the window function within
$-T{\leq}t^\prime{\leq}-t_\eta$:
\begin{equation}
\label{five}
  g(-t')
    =  \frac{(L/2)}
            {1+\frac{(L-\eta)}{\eta}\frac{(T+t')}{(T-t_\eta)}}
       \; ,
\end{equation}
which also satisfies the boundary conditions $g(T){=}L/2$ and 
$g(t_\eta){=}\eta/2$. So far, no specification of 
$g(-t')$ for $-t_\eta\leq t'\leq 0$ has been given; since $t_\eta{\ll}T$
should hold on physical grounds, the simplest choice would be 
$t_\eta=0$.

The construction proposed in Eqs.\ (\ref{two})-(\ref{five}) guarantees
that the one-point statistics of the dynamical RMCM is identical to its
geometrical counterpart. In order to qualify for a complete dynamical
generalisation, it is not only the one-point statistics, but the
$n$-point statistics in general, which should match. Hence, we now 
consider the equal-time two-point correlator
\begin{eqnarray}
\label{six}
  C_{n_1,n_2}(l{=}x_2{-}x_1{\geq}\eta)
    &=&  \frac{ \left\langle
                \varepsilon^{n_1}(x_1,t) \, \varepsilon^{n_2}(x_2,t)
                \right\rangle }
              { \left\langle \varepsilon^{n_1}(x_1,t) \right\rangle 
                \left\langle \varepsilon^{n_2}(x_2,t) \right\rangle}
         \nonumber \\
    &=&  \frac{ \left\langle D(l)^{n_1+n_2} \right\rangle}
              { \left\langle D(l)^{n_1} \right\rangle
                \left\langle D(l)^{n_2} \right\rangle }
         \; .
\end{eqnarray}
The correlation between the two points with distance $l{<}L$ stems from
the overlap region of the two index functions $f(x_1,t)$ and 
$f(x_2{=}x_1+l,t)$; see Fig.\ 2a. This explains the second step of
(\ref{six}), where
\begin{equation}
\label{seven}
  D(l{\geq}\eta)
    =  \exp\left(
       \int_{-T}^{-g^{-1}(l/2)} dt'
       \int_{l-g(-t')}^{g(-t')} dx' \gamma(x',t')
       \right)
\end{equation}
represents the contribution from the overlap region. The contributions 
from the non-overlapping regions are statistically independent, hence 
factorise and cancel. Introducing the spatio-temporal overlap volume
\begin{eqnarray}
\label{eight}
  V(l{\geq}\eta)
    &=&  \int_{-T}^{-g^{-1}(l/2)} dt'
         \int_{l-g(-t')}^{g(-t')} dx' 
         \nonumber \\
    &=&  \frac{\eta{L}(T-t_\eta)}{(L-\eta)}
         \ln\left( \frac{L}{l} \right)
         - \frac{\eta(T-t_\eta)}{(L-\eta)} (L-l)
\end{eqnarray}
and employing basic properties of stable distributions \cite{SAM94}, 
the expectation of the $n$-th power of expression (\ref{seven}) can be 
transformed into
\begin{equation}
\label{nine}
  \left\langle D(l)^n \right\rangle
    = \exp\left( \frac{\sigma^\alpha}{\cos\frac{\pi\alpha}{2}} 
                 V(l) (n-n^\alpha) \right)
         \; .
\end{equation}
Defining the multifractal scaling exponents
$\tau(n)=\tau(2)(n{-}n^\alpha)/(2{-}2^\alpha)$ with
$\tau(2)=(\sigma^\alpha/\cos\frac{\pi\alpha}{2})(2{-}2^\alpha)
         \eta L(T{-}t_\eta)/(L{-}\eta)$ 
as well as $\tau[n_1,n_2]=\tau(n_1+n_2)-\tau(n_1)-\tau(n_2)$,
insertion of (\ref{nine}) into (\ref{six}) leads to the final 
expression for the two-point correlator:
\begin{equation}
\label{ten}
  C_{n_1,n_2}(l)
    =  \left( \frac{L}{l} \right)^{\tau[n_1,n_2]}
       \exp\left( -\tau[n_1,n_2]\left(1-\frac{l}{L}\right) \right)
       \; .
\end{equation}
It reveals multiscaling behaviour for $\eta{<}l{\ll}L$, which 
shows that the equal-time two-point statistics of the dynamical
RMCM in $1{+}1$ dimensions is completely analogous to the 
findings of the geometrical RMCM in $1{+}0$ dimensions.

In view of this analogy, the large-scale deviation appearing in the last 
expression is reminiscent of similar recent findings on scaling functions 
within the geometrical RMCM\cite{SCH00}. With the introduction of a 
suitably tuned large-scale fluctuation this large-scale deviation can be 
cancelled. In our present context this implies a small extension of the 
window function (\ref{five}) beyond the time interval
$-T{\leq}t^\prime{\leq}0$: $g(T{\leq}{-}t'{\leq}T{+}\Delta{T}) = L/2$;
consult again Fig.\ 1. It is then straightforward to show that for 
$\Delta{T} = (T-t_\eta)\eta/(L-\eta)$, which is much smaller than the 
correlation time $T$, the two-point correlator exactly becomes
$C_{n_1,n_2}(l) = (L/l)^{\tau[n_1,n_2]}$ for all $\eta{\leq}l{\leq}L$.

Besides equal-time two-point correlations temporal two-point 
correlations are also of interest. These correlations arise due to a 
modified spatio-temporal overlap volume as illustrated in Fig.\ 2b.
A straightforward calculation, analogous to (\ref{seven})-(\ref{nine}), 
yields
\begin{eqnarray}
\label{eleven}
  C_{n_1,n_2}(t=t_2-t_1)
    &=&  \frac{ \left\langle
                \varepsilon^{n_1}(x,t_1) \varepsilon^{n_2}(x,t_2)
                \right\rangle }
              { \left\langle \varepsilon^{n_1}(x,t_1) \right\rangle 
                \left\langle \varepsilon^{n_2}(x,t_2) \right\rangle}
         \nonumber \\
    &=&  \left(
         \frac{t}{T} + \frac{\eta}{L} \left( 1-\frac{t}{T} \right)
         \right)^{-\tau[n_1,n_2]}
         \nonumber \\
    &\approx&  
         \left( \frac{T}{t} \right)^{\tau[n_1,n_2]}
         \; ;
\end{eqnarray}
for simplicity, the parameter of the window function (\ref{five}) has 
been set $t_\eta=0$. The last step, only holding for $0{\ll}t{<}T$ and 
$\eta{\ll}L$, reveals that the temporal two-point correlator comes
with scaling exponents identical to those of the equal-time two-point
correlators.

Within this dynamical RMCM investigations on several other observables 
are nearly as straightforward \cite{SCH01}. Here we mention only
spatio-temporal $(n{\geq}2)$-point correlations in connection with 
fusion rules, and moments and multiplier distributions based on 
coarse-grained field amplitudes. All of these observables are found
to be in good agreement with experimental results derived from the 
surrogate energy dissipation field of fully developed turbulence.

The dynamical RMCM, presented here, is a generalisation of the 
geometrical RMCM. By construction it is continuous and homogeneous, 
does not make use of a discrete hierarchy of scales and stochastically 
evolves a positive-valued field in one space and one time dimension. 
Several generalisations of this new model immediately come to mind:
stochastic evolution in $n{+}1$ dimensions with the optional inclusion
of spatial anisotropy, discretisation of space-time into smallest cells
to model dissipation and deviation from log-stability, and dynamical 
RMCM for vector fields to model the turbulent velocity field. Work
in these directions is in progress.

\vspace*{2cm}
The authors acknowledge fruitful discussions with Jahanshah Davoudi. 
This work has been supported in parts by DAAD and by BCPL in the 
framework of the European Community-Access to Research
Infrastructure action of the Improving Human Potential Program.


\newpage
\mbox{ }\hspace*{-15mm}
\parbox[t]{65mm}{  \psfig{file=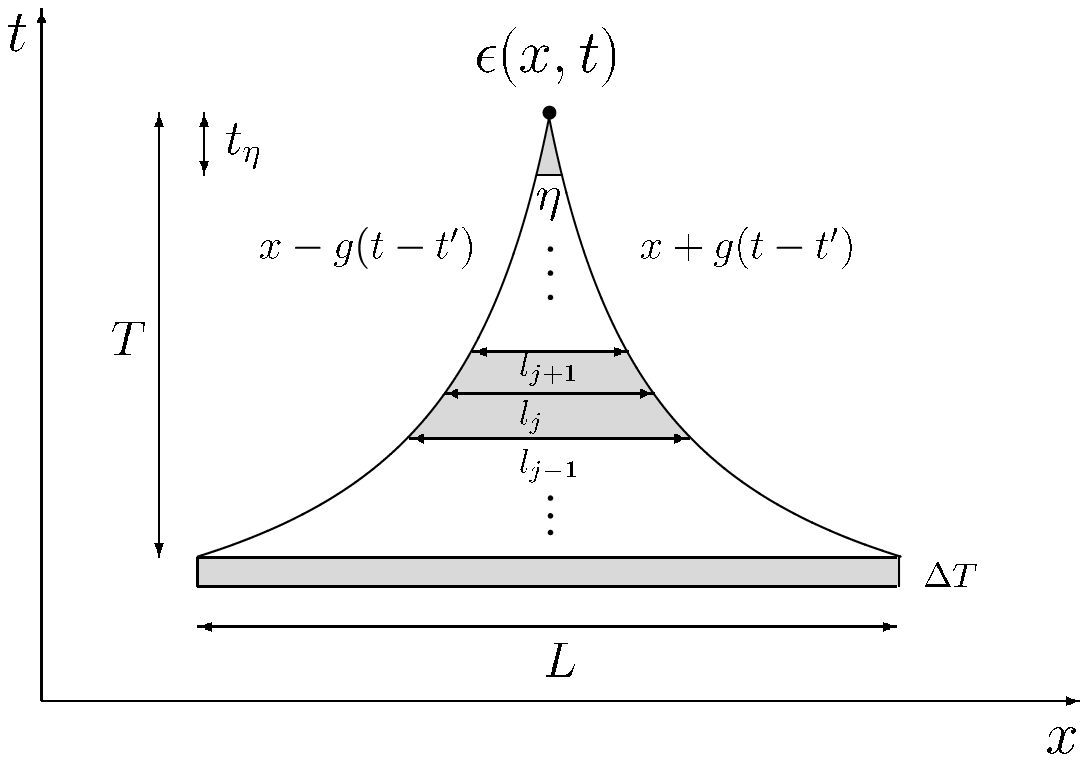,height=110mm}   }
\par\ \\
Figure 1:
Illustration of the causal index function (\ref{three}) used to 
construct the positive-valued multifractal field $\varepsilon(x,t)$. 

\newpage
\mbox{ }\hspace*{-15mm}
\parbox[t]{65mm}{  \psfig{file=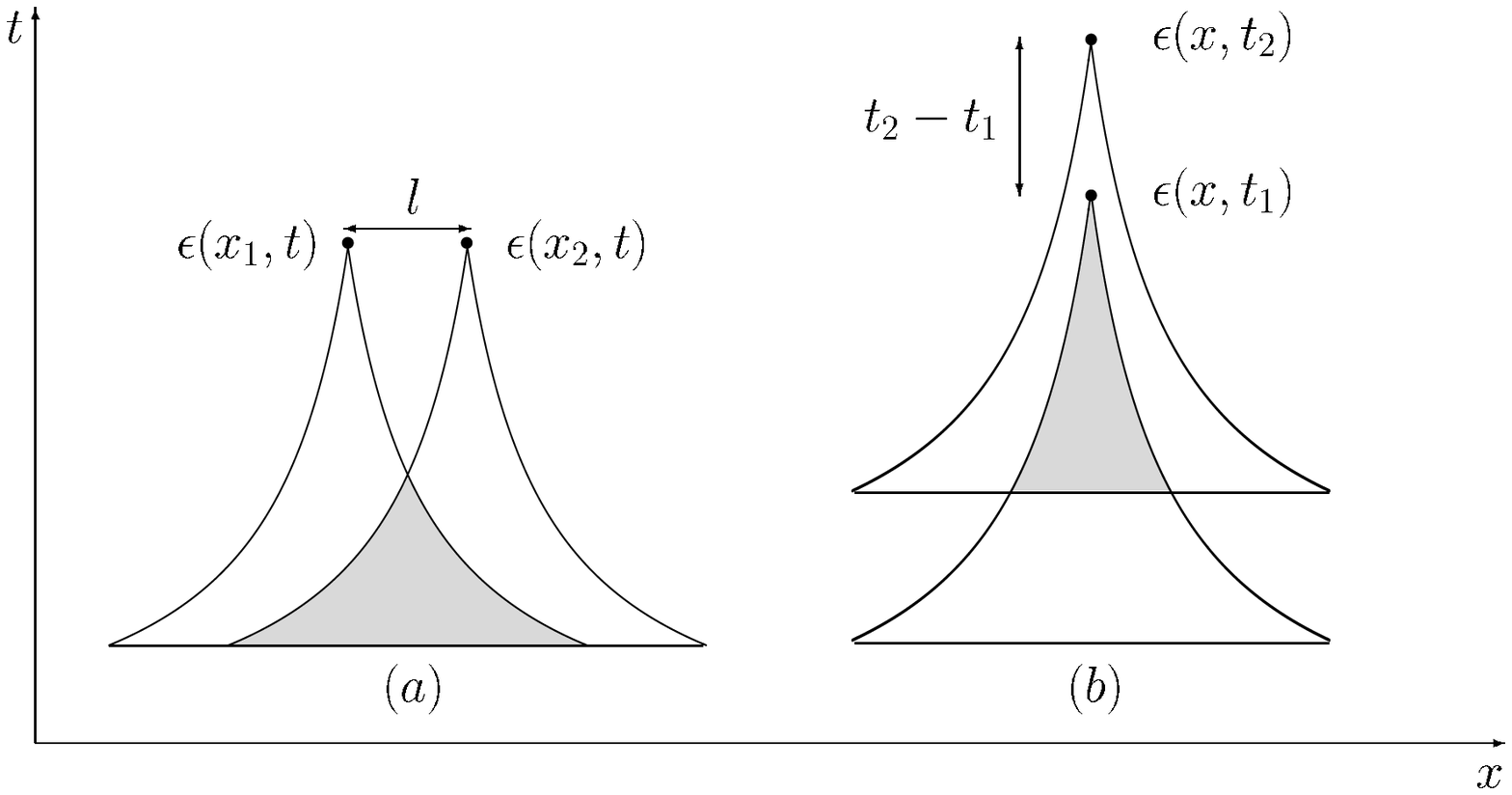,height=85mm}   }
\par\ \\
Figure 2:
Spatio-temporal overlap volumes (shaded) producing the correlation for
the (a) equal-time and (b) temporal two-point correlator.

\end{document}